\documentstyle[psfig]{l-aa}             
 
\begin{document}

\def\diff{\partial} 
\def\d{ {\rm d} }   
\def\vec#1{\bf #1}  

\def\n{\noindent}
\let\speciali =\n
\def\ol{\overline}
\let\specialii =\ol
\def\O{\Omega}
\let\specialiii =\O
\def\wt{\widetilde}
\let\specialiv =\wt
\def\wh{\widehat}
\let\specialv =\wh
\def\f{f(\theta)}
\let\specialvi =\f
\def\k{\rho \chi}
\let\specialvii =\k
\def\en{\rho \varepsilon}
\let\specialviii =\en


\newcommand{\beq}{\begin{equation}}
\newcommand{\beqa}{\begin{eqnarray*}}
\newcommand{\beqan}{\begin{eqnarray}}
\newcommand{\greq}{\begin{equation}\left\{ \begin{array}{l}}
\newcommand{\eeq}{\end{equation}}
\newcommand{\eeqa}{\end{eqnarray*}}
\newcommand{\eeqan}{\end{eqnarray}}
\newcommand{\hsp}{ \hspace{.5cm} }
\newcommand{\noi}{ \noindent }
\newcommand{\ssi}{ \Longleftrightarrow}
\newcommand{\lp}{ \left(}
\newcommand{\rp}{ \right)}
\newcommand{\lc}{ \left[}
\newcommand{\rc}{ \right]}
\newcommand{\dzeta}{\zeta}
\newcommand{\khi}{\chi}
\newcommand{\YL}{ Y^m_l }
\newcommand{\PL}{ P^m_l }
\newcommand{\eps}{\varepsilon}
\newcommand{\na}{ \vec{\nabla} }
\newcommand{\intsur}{ \int_{(S)}\! }
\newcommand{\intvol}{ \int_{(V)}\! }
\newcommand{\cth}{ \cos\theta }
\newcommand{\sth}{ \sin\theta }
\newcommand{\dOm}{d\Omega}
\newcommand{\vvphi}{\vec{v}_{\phi}}
\newcommand{\demi}{\frac{1}{2}}
\newcommand{\omb}{\overline{\omega}}
\newcommand{\rhobar}{\overline{\rho}}
\newcommand{\ephi}{\vec{e}_\phi}
\newcommand{\ddr}[1]{\frac{{\rm d}  #1}{{\rm d} r}}
\newcommand{\dr}[1]{\frac{\partial  #1}{\partial r}}
\newcommand{\dt}[1]{\frac{\partial  #1}{\partial t}}
\newcommand{\dnt}[1]{\frac{{\rm d}  #1}{{\rm d}t}}
\newcommand{\dnx}[1]{\frac{{\rm d}  #1}{{\rm d}x}}
\newcommand{\dny}[1]{\frac{{\rm d}  #1}{{\rm d}y}}
\newcommand{\dnz}[1]{\frac{{\rm d}  #1}{{\rm d}z}}
\newcommand{\Dt}[1]{\frac{D #1}{D t}}
\newcommand{\dx}[1]{\frac{\partial  #1}{\partial x}}
\newcommand{\dy}[1]{\frac{\partial  #1}{\partial y}}
\newcommand{\dz}[1]{\frac{\partial  #1}{\partial z}}
\newcommand{\dtheta}[1]{\frac{\partial  #1}{\partial \theta}}
\newcommand{\drtheta}[1]{\frac{1}{r}\frac{\partial  #1}{\partial\theta}}
\newcommand{\ddrr}[1]{\frac{\partial^2  #1}{\partial r^2}}
\newcommand{\ddtt}[1]{\frac{\partial^2  #1}{\partial t^2}}
\newcommand{\drr}{\frac{\partial}{\partial r}}
\newcommand{\dxx}{\frac{\partial}{\partial x}}
\newcommand{\dtt}{\frac{\partial}{\partial t}}
\newcommand{\ksi}{\xi}


\bibliographystyle{plain}

\thesaurus{}

\title{Rotational mixing in early-type stars: the main-sequence evolution of a 
9 M$_\odot$ star.}
\author{Suzanne Talon\inst{1}, Jean-Paul Zahn\inst{1}, Andr\'e Maeder\inst{2} 
and Georges Meynet\inst{2}}
\offprints{S. Talon }

\institute{
D\'epartement d'Astrophysique Stellaire et Galactique, Observatoire de
 Paris, Section de Meudon, 92195 Meudon, France
\and
Observatoire de Gen\`eve, CH-1290, Sauverny, Switzerland \\
e-mail : talon@obspm.fr, zahn@obspm.fr, maeder@scsun.unige.ch,
meynet@scsun.unige.ch}


\maketitle
\markboth{S. Talon et al.: Rotational mixing in early-type stars }{}

\begin{abstract}
We describe the main-sequence evolution of a rotating 9 M$_\odot$ star.
Its interior rotation profile is determined by the redistribution of
angular momentum through the meridian circulation and through the shear
turbulence generated by the differential rotation; the possible effect of
internal waves is neglected.  We examine the mixing of chemicals produced by
the same internal motions.  Our modelization is based on the set of equations
established by Zahn (1992) and completed in Matias, Talon \& Zahn (1996). 
Our calculations show that the amount of mixing
associated with a typical rotation velocity of $\sim 100~{\rm km \, s^{-1}}$
yields stellar models whose global parameters are very similar to those
obtained with the moderate overshooting ($d/H_P \simeq 0.2$) which has been
invoked until now to fit the observations. 
Fast rotation ($\sim 300~{\rm km \, s^{-1}}$) leads to significant changes of
the C/N and O/N surface ratios, but the abundance of He is barely increased. 
The modifications of the internal composition profile due to such
rotational mixing will certainly affect the post--main-sequence evolution.

\end{abstract}

\section{Introduction}

There are many reasons to suspect that the amount of mixing
presently included in stellar models, in the form of overshooting or
semi-convection, is insufficient to explain some observational facts (Maeder
1995). In particular, Herrero et al. (1992) found a strong correlation between
He enrichment and surface velocity in O-type stars, which indicates that there
is probably a link between rotation and mixing.

Indeed it is well known, since the early work of Eddington (1925)
and Vogt (1925), that the meridional currents arising in a rotating star may
cause internal mixing. 
But until recently, the only case considered seriously was that of
uniformly rotating stars, and it has been proved by Mestel (1953) that in such
stars the stabilizing effect of the composition gradients, which arise from
hydrogen burning, is so strong that even the fastest rotators are unlikely to
be mixed.

The situation is different, however, if one allows the circulation to modify
the rotation profile, as has been pointed out by Zahn (1992; Paper I). He
invoked the anisotropic turbulence observed in stratified fluids, with much
stronger transport in the horizontal than in the vertical direction, to reduce
the problem to one dimension, with the angular velocity $\Omega$ depending
only on depth. Then the evolution of $\Omega$ obeys a ``hyper-diffusion''
equation, involving the fourth derivative of $\Omega$, and in massive
main-sequence stars the rotation profile tends to adjust such as to minimize
the flux of angular momentum, as had been anticipated by Busse (1981).  
Provided there is no interference from other transport processes, the residual
turbulence associated with that differential rotation is the main cause of
mixing in the radiative interior, and this is true even in the regions of
strong composition gradient which arise near the convective core, thanks to the
homogenizing action of the anisotropic turbulence, as shown recently by Talon
\& Zahn (1996).

Other groups have worked on rotational mixing in early-type stars,
and they have clearly shown that such mixing may modify the surface
abundances of He, C, N and O (Langer 1991, 1992; Denissenkov 1994;
Eryurt et al. 1994).
However, they have used an older prescription based largely on dimensional
arguments, which links that mixing to the rotation velocity (Zahn 1983). 
Meynet \& Maeder (1996) performed similar calculations, but 
they took into account differential rotation, and calculated
the diffusion related to it, assumimg that the rotation profile is determined
by the local conservation of angular momentum.
The improvement brought by the present paper is that here the internal 
rotation evolves for the first time consistently under the combined action of
the meridian circulation and the shear turbulence.
Furthermore, as explained in Zahn (1992) and in Matias et al. (1996),
our treatment of rotational mixing involves only one free parameter,
which we hope to calibrate in the near future using fast rotators.

In this paper our goal is more modest, for we shall not consider stars as 
massive as those observed by Herrero et al. 
The main reason is that we do not want our 
results to dependent too drastically on the prescription employed to treat
semi-convection, which significantly affects the structure
of the inner portion of O-stars.
We choose to describe the evolution of a 9 M$_\odot$ star, and intend to
compare our results with the observations used to calibrate overshooting in
moderate mass stars.

We begin by presenting our stellar models and the equations governing the
transport of angular momentum and of chemical elements (\S 2). The results for
a hypothetical, stationary model are discussed in (\S 3), and those for the
full main-sequence evolution in (\S 4).


\section{Stellar models with rotation}

\subsection{Input physics}

Our models were built with the Geneva
stellar evolution code, and they were evolved starting
from a homogeneous composition at the zero-age main-sequence (ZAMS).

The equation of state includes partial ionization
close to the surface of the star.

We used the OPAL radiative opacities (Iglesias et al. 1992)
which include the spin-orbit interactions for Fe.
The relative metal abundances are based on Grevesse (1991),
consistently with the opacity used. 
Note however that this
consistency will be lost in the course of evolution, since mixing
and nuclear processing
modify those relative abundances. 
The tables are completed for the lower
temperatures (below 10~000K) by those given by Kurucz (1991).

Hydrogen burning proceeds through the p-p chains and the
CNO cycle. The reaction rates are as given by Caughlan \&
Fowler (1988) with the correction 
by Landr\'e et al. (1990) for the reactions $^{17}{\rm O}({\rm p}, 
\gamma )^{18}{\rm F}$ and $^{17}{\rm O}({\rm p},\alpha)^{14}{\rm N}$.

For the mixing length parameter $\alpha = l/H_P$ we take the value 1.6,
drawn from the solar calibration with the same code (cf. Schaller et
al. 1992).

No allowance will be made here for convective penetration, which is still a
matter of debate. 
Most theoretical predictions (Roxburgh 1978, 1989; Zahn 1991) tend to
overestimate this penetration, and the reason may be that they do not
account for the inhibiting effect of rotation, which has been clearly
demonstrated in the recent numerical simulations performed by Julien et al. (1996).

One important observational constraint for models of OB stars is the width of
the main-sequence, and it can be accounted for by an overshooting parameter
$d/H_p \simeq 0.2$  (cf. Stothers \& Chin 1991).
However, this value characterizes in fact the size of the well mixed core
of the star, and that mixing may be due either to convective penetration or to any
other mechanism. 
In this study, we shall examine whether this core extension
may be attributed to rotational mixing, as described by Zahn's formalism 
(cf. Paper I).
Therefore, we assume that convective overshooting is rather weak, just enough
to yield a finite value of the subadiabatic gradient ($\nabla _{\rm ad} - 
\nabla _{\rm rad}$), in order to prevent the divergence of the circulation
velocity at the boundary of the convective core.

Mass loss is also included, following the
empirical relation given by de Jager et al. (1988), which
is valid for O to M main-sequence stars.

Furthermore, the equilibrium structure of our models 
includes the mean centrifugal force (averaged over an isobar).  The
hydrostatic equation reads:  
\beq
\frac{1}{\rho} \ddr{P} = - \frac{Gm}{r^2} + \frac{1}{2}\left<\sin ^2 \theta \right> \O ^2 r 
\eeq 
where we use standard notations for the pressure ($P$), density ($\rho$),
radius ($r$), mass ($m$), gravitational constant ($G$), colatitude ($\theta$),
angular velocity ($\O$) and where the horizontal average of $f$ is defined as
\beq
\left<f \right> = \frac{\int_0^\pi f \sin \theta {\rm d}\theta}
{\int_0^\pi \sin \theta {\rm d}\theta}.   \label{hmean}
\eeq
We shall not refine here the implementation of the centrifugal force, since
it would have little effect on mixing; for a more rigorous treatment, we refer
the reader to Meynet \& Maeder (1996), where they adapt the Kippenhahn \&
Thomas (1970) method 
to incorporate the hydrostatic effects of rotation in 
one dimension to the case of shellular rotation.

\subsection{Transport equations}

In this paper, we shall consider only the transport of matter and angular
momentum due to meridional circulation and shear turbulence. Microscopic
diffusion has negligible effect in early-type stars, and it will be ignored
here. The possible redistribution of angular momentum through
internal waves is also neglected, although it has been shown to play an
important role in the Sun (Kumar \& Quataert 1996; Zahn, Talon \& Matias 1996);
we shall revisit the problem once we dispose of a reliable prescription for
this transport in a rapidly rotating star.

The equations controlling the
evolution of the rotation profile through meridional advection and
turbulent diffusion have been presented in Paper I and in Matias et al.
(1996).  Let us recall that the main assumption in this description
is the strong anisotropy of the turbulence, whose horizontal transport
enforces a state of shellular rotation $\O = \O(P)$, so that all variables
may be expressed with respect to the isobars as
\beqan 
f(P,\theta) &=& \left< f(P,\theta) \right> + 
\widetilde{f}(P) P_2(\cos~\theta) \\ \nonumber
&=& f(P) + \widetilde{f}(P) P_2(\cos~\theta). 
\eeqan

After performing the suitable horizontal means, the transport
equation for angular momentum is cast into
\beq
\rho \dtt \lc r^2 {\O}\rc = \frac{1}{5 r^2} \drr \lc \rho r^4 {\O}
U \rc + \frac{1}{ r^2} \drr \lc \rho \nu_v r^4 \dr{\O} \rc
\label{ev_omega}
\eeq
(cf. Eq. (2.7) of Paper I), where $\nu_v$ is the vertical (turbulent) viscosity
and $U$ the amplitude of the vertical circulation speed $u(P, \theta)=U(P)
P_2(\cos \theta)$:
\beq
U(P)  = {L \over m g} \left( {P \over C_P \rho T} \right)
 {1 \over \nabla _{\rm ad} - \nabla} \lc E_{\O} + E_{\mu} \rc ,
\label{meru}
\eeq
with $L$ being the luminosity, $g$ the local gravity, $C_P$ the specific heat,
$T$ the temperature.  The function $E_{\O}$ depends on the velocity profile
\beqan
E_{\O} &=&
2\, \left[
1 - {{\O^2} \over 2 \pi G  \rho}
- {{\varepsilon} \over \varepsilon _m} \right]
{\wt g \over  g}  \nonumber \\
& -  &{\rho _m \over  \rho} \, \left[
{r \over 3} \, {\diff \over \diff r}
 \left(  H_{\rm T} {\diff \Theta \over \diff r}
- \chi _{\rm T} \, \Theta \right) - 2 {H_{\rm T} \over r} \Theta
+ {2 \over 3} \Theta \right] \nonumber \\
& -  & { \varepsilon
\over \varepsilon _m} \,
\left[ H_{\rm T} {\diff \Theta \over \diff r}
+ (\varepsilon _{\rm T} - \chi _{\rm T}) \, \Theta \right]
  \label{eomeg} 
\eeqan
and $E_{\mu}$, on the chemical inhomogeneities along isobars
        \beqan
E_{\mu} &=&
   {\rho _m \over  \rho} \, \left[
{r \over 3} \, {\diff \over \diff r}
 \left(  H_{\rm T} {\diff \Lambda \over \diff r}
- (\chi _{\rm T} +1) \, \Lambda  - \sum _i \chi _{c_i}
\frac{\wt c_i}{c_i} \right) \right.\nonumber \\
&-&  \left. 2 {H_{\rm T} \over r} \, \Lambda  \right]
+  { \varepsilon
\over  \varepsilon _m}\,
\left[ H_{\rm T} {\diff \Lambda \over \diff r}
+ (\varepsilon _{\rm T}
- \chi _{\rm T} - 1)
\, \Lambda \right. \label{emu} \\
&+& \left. \sum _i \lp \varepsilon _{c_i} - \chi _{c_i} \rp \frac{\wt c_i}{c_i}
\right] \nonumber
        \eeqan
(cf. Matias et al. 1996).
Here $\varepsilon$ is the nuclear energy
generation rate, $H_T$ the temperature scale height, $\chi$ the radiative
conductivity, and $c_i$ the concentration of a chemical species of
mass number $i$; the subscript $m$ designates the mean value of the considered
variable over the isobar, and other subscripts refer to logarithmic
derivatives. $\Theta=r^2/3g \partial \O^2/\partial r$ is a measure of the
differential  rotation in the vertical direction and $\Lambda=
\widetilde{\mu}/\mu$ is the variation amplitude of the mean molecular weight
along an isobar. This amplitude is derived from the horizontal variation of
the concentration $\wt c_i$ of each species, which results from the competition
between horizontal diffusion and vertical advection:  
       \beq
\wt {c _i} = - {r^2 U \over 6 \, D_h} \,
{\diff {c _i} \over \diff r} \, .
\label{fluctc}
       \eeq
The consequence is that the vertical transport due to the meridian
circulation is turned into a diffusion, with an effective diffusivity given by
(Chaboyer \& Zahn 1992):
       \beq
D_{\rm eff} = {|r U(r)|^2 \over 30 \, D_h} \; .
\label{deff}
       \eeq

Thus the transport of a given chemical element varies with time according to
       \beq
\rho {\partial c_i \over \partial t}   =  \dot{c_i} +
 {1 \over r^2} {\partial \over \partial r}
\left[ r^2 \rho \left(D_{\rm eff} + D_v \right)
{\partial c_i \over \partial r} \right]  \; , 
\label{trac}
        \eeq
where the nuclear production/destruction rate $\dot{c_i}$ is
taken from the stellar structure code.

In this study, the only contribution retained in the {\it vertical} turbulent
diffusivity ($D_v$) is that of the vertical shear, and we assume that this
diffusivity equals the turbulent viscosity. We use the prescription given by
Talon \& Zahn (1996): 
\beq
\nu_v = D_v = \frac{8 Ri_c }{5} \, \frac {(r \, {\rm d}\O/{\rm d}r)^2}
{N^2_T/(K+D_h) + N^2_{\mu}/D_h} ,
 \label{TZ}
\eeq
which takes into account the homogenizing effect of the horizontal diffusion
($D_h$) on the restoring force produced by the $\mu$-gradient.
Here, the Brunt-V\"{a}is\"{a}l\"{a} frequency has been split into
$N^2_T = \frac{g \delta}{H_P} \lp \nabla _{\rm ad} - \nabla \rp$
and
$N^2_\mu = \frac{g\varphi}{H_P}   \nabla _\mu$,
$K$ is the radiative diffusivity and $Ri_c
\approx 1/4$ is the critical Richardson number.

It remains to evaluate the magnitude of the horizontal
diffusion ($D_h$), which we are unable to derive from first principles.
We will use a parametric relation established as follows.
Since the horizontal shear is sustained by the advection of
momentum, $D_h$ must be related to the
circulation velocity $U$ in order to keep the differential
rotation in latitude below a certain level. This leads to
\beq
D_h = \frac{rU}{C_h} \lc \frac{1}{3} \frac{{\rm d} \ln \rho r^2 U}
{{\rm d} \ln r} - \frac{1}{2} \frac{{\rm d} \ln r^2 \O}{{\rm d} \ln r} \rc,
\eeq
where $C_h$ is an unknown parameter of order unity (see Zahn 1992).

\subsection{Numerical technique}

At each time step, the stellar evolution code is used to compute the mean
structure of the model. Then, the meridional circulation is calculated,
with the induced diffusion. The new rotation profile, as well as
the new chemical abundances, are then implemented back into the
evolutionary code, and we iterate until we reach the solution.

The fourth order equation governing the evolution of the rotation
profile is solved by a relaxation method based on finite
differences, as in the Henyey method employed in the stellar evolution code
(cf. Henyey et al. 1964).

The diffusion equation is solved from the center of the star up to
the surface, with a finite element method, using a semi-implicit
temporal integration. The convective zone is homogenized by diffusion,
using an arbitrary (large) coefficient of $D=10^{14} {\rm cm~ s^{-1}}$.

\section{Asymptotic regime}

\begin{figure}[t]     
\centerline{
\psfig{figure=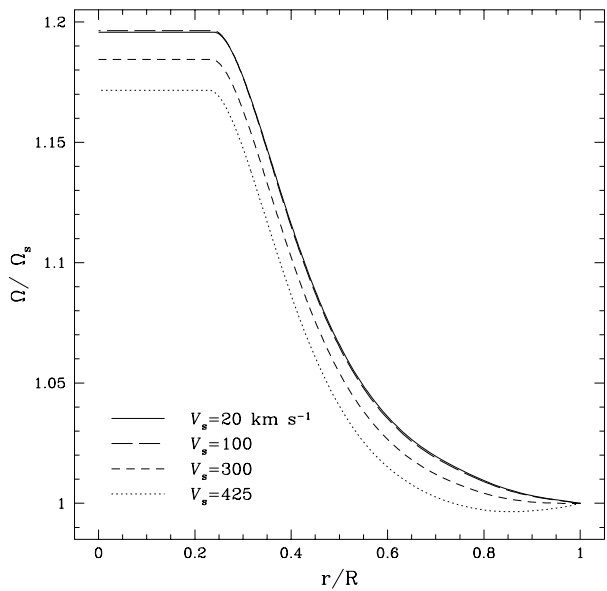,height=8.4cm}
}
\caption[]{Normalized rotation profiles in the asymptotic regime, for a $9
M_{\odot}$ star. The solid line represents the star with a surface velocity of
$20~{\rm km \, s^{-1}}$, the long dashed line the same with a velocity
of $100~{\rm km \, s^{-1}}$, the short dashed line with a
velocity of $300~{\rm km \, s^{-1}}$ and the dotted line, the fastest rotator
with a surface velocity of $425~{\rm km \, s^{-1}}$.}
\end{figure}

\begin{figure}[t]  
\centerline{
\psfig{figure=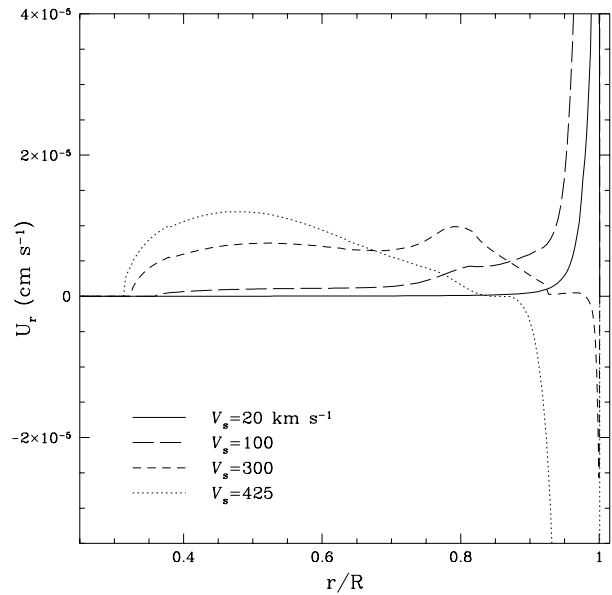,height=8.4cm}
}
\caption[]{Circulation velocity in the radial direction $U(r)$, for the
asymptotic regime. (Line styles are defined as in Fig.~1.)
}
\end{figure}

\begin{figure}[t]  
\centerline{
\psfig{figure=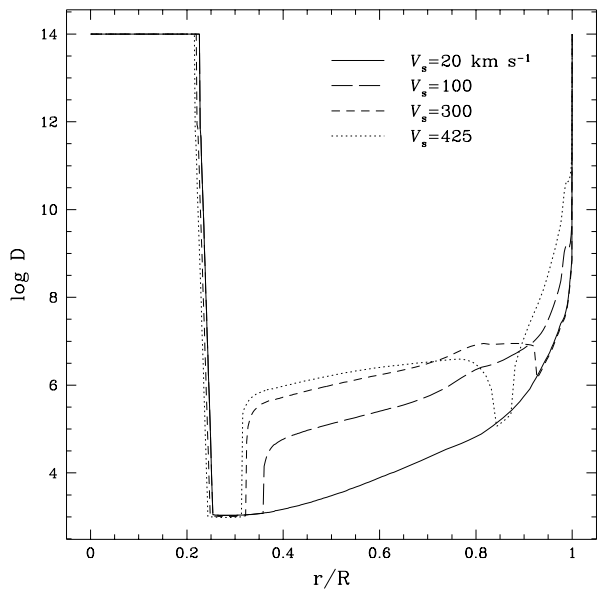,height=8.4cm}
}
\caption[]{Diffusion coefficients in the asymptotic regime. (Line styles
are defined as in Fig.~1.)
}
\end{figure}

\begin{figure}[t]  
\centerline{
\psfig{figure=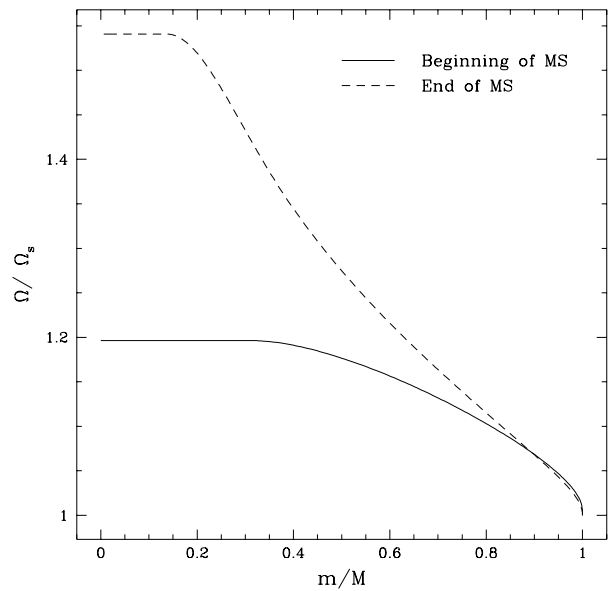,height=8.4cm}
}
\caption[]{Rotation profile required by thermal equilibrium;
comparison between a ZAMS model and a TAMS model.
}
\end{figure}

\begin{figure}[t]  
\centerline{
\psfig{figure=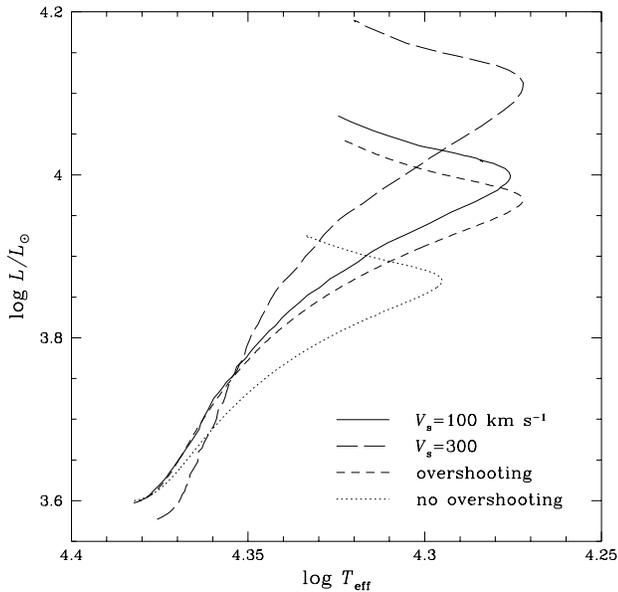,height=8.4cm}
}
\caption[]{The evolution of a rotating star of 9 $M_\odot$ in the HR
diagram. The solid line tracks the evolution of a star
with an initial surface velocity of $100~{\rm km \, s^{-1}}$, the long dashed
line that of a star with an initial surface velocity of
$300~{\rm km \, s^{-1}}$. For reference, the short dashes show the evolution
of the same star with an overshooting of $d/H_p = 0.2$
and the dotted line the star without any overshooting.
}
\end{figure}

\begin{figure}[t]  
\centerline{
\psfig{figure=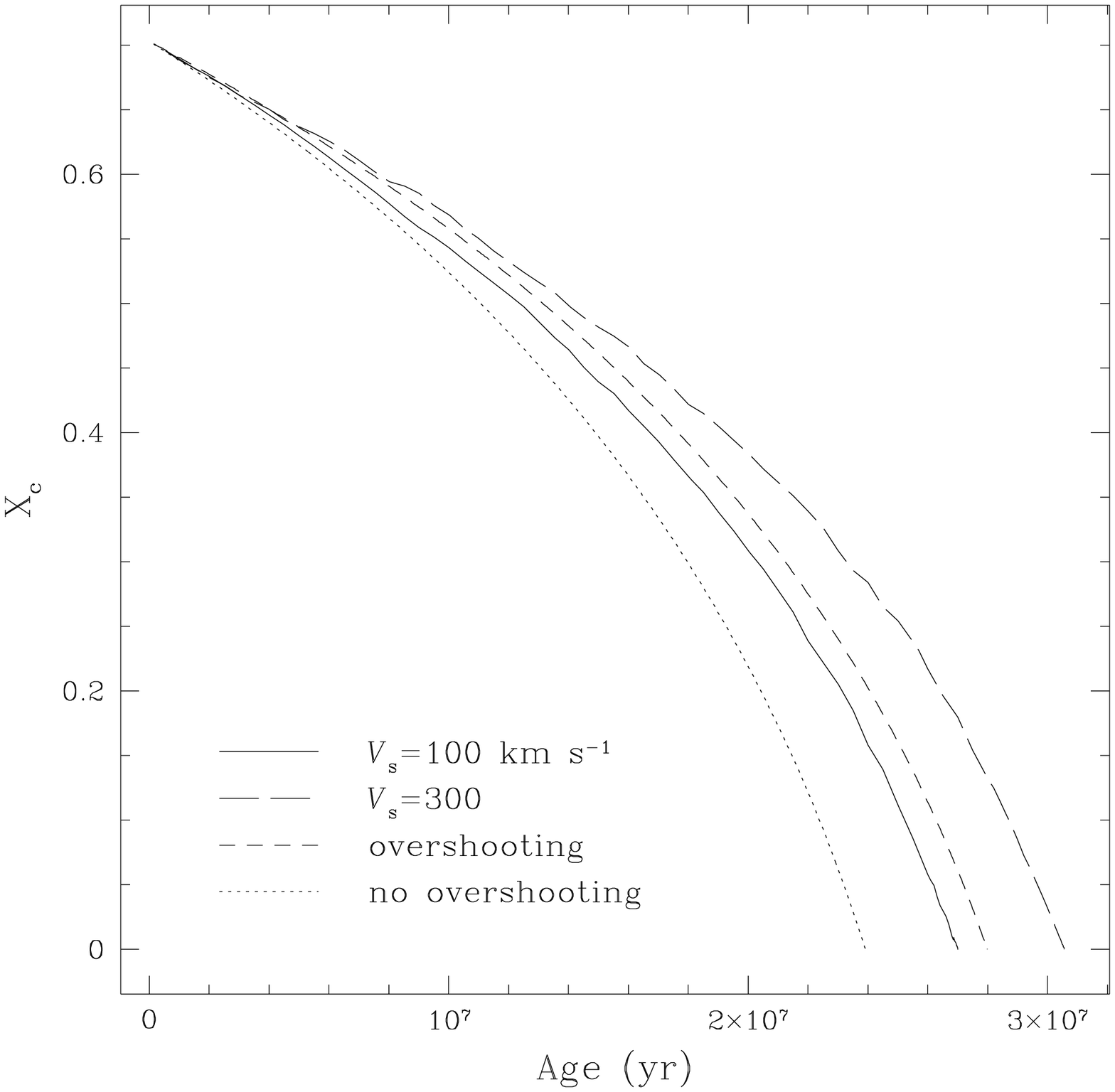,height=8.4cm}
}
\caption[]{Evolution of the central abundance of hydrogen.
(Line styles
are defined as in Fig.~5.)
}
\end{figure}

When the star does not lose angular momentum, which to first approximation
is the case of massive stars, the internal rotation rate tends to a profile
which cancels the flux of angular momentum in the radiative interior. If the
star was not evolving, the flow would settle into a stationary
regime, in which the advection of angular momentum by the circulation exactly
balances the diffusive transport:
\beq -\frac{1}{5} U(P) = \nu_v
\ddr{\ln \O}. \label{reg_ass}
\eeq 
Although this state will never be achieved in a real star, it is worth
considering it for the purpose of comparison with other results.

\subsection{Asymptotic circulation profile in realistic models}

We calculated the stationary regime obtained for a ZAMS
star of $9 M_{\odot}$. For this, we solved the p.d.e. (fourth order in
space, first order in time) governing the evolution of
angular momentum (Eq.~{\ref{ev_omega}),
starting from solid body rotation, until we reached a stationary
regime.
Urpin et al. (1996) performed a similar
study, for $20 M_{\odot}$ models. However, they used a procedure slightly
different from ours, searching directly the solutions of the third
order o.d.e. (\ref{reg_ass}).
We shall compare our results to theirs when possible.

In Fig.~1, we show the rotation profiles obtained
for four different surface velocities
($20$, $100$, $300$ and $425~{\rm km \, s^{-1}}$).
Figure~2 displays the corresponding radial velocities of the meridian
circulation.

We observe that 
the normalized profile is independent of the rotation velocity
in the inner portion of the star, 
whereas the situation is quite different close to the surface.
The reason is that $\nu_v$ scales as $\O^2$, and so
does also $E_\O$, through  $\wt g / g$ and $\Theta$, as long as the next order
term $-\O^2/2 \pi G \rho$ remains negligible compared to unity, which is the
case in the deep interior (cf. Eq. \ref{eomeg}).
Then Eq.~(\ref{reg_ass}) is insensitive to the magnitude of $\O$, and it can
only determine the shape of the rotation profile. 
That higher order term, which we owe to Gratton (1945) and \"Opik (1951),
dominates however near the surface, and it renders $E_\O$ nonlinear in $\O^2$,
thus modifying the rotation profile. For rapid rotation, the radial derivative
of $\O$ becomes positive near the surface, imposing a reversal of the diffusive
flux and therefore also of the radial component of the circulation velocity,
as can be seen in Fig.~2.

Urpin et al. reach similar conclusions. They obtain a
differential rotation between center and surface of $\simeq 1.15$,
which is comparable to what we get. Their fastest rotator (\#4) corresponds
to our  model with $300~ {\rm km \, s^{-1}}$. Its circulation
velocity  becomes marginally negative near the surface, as for ours.
However, the velocities they obtain for the inner part of their models
($\sim 50~ {\rm cm \, s^{-1}}$) imply viscosities of order
$\sim 10^{12}~ {\rm cm^2\, s^{-1}}$, values that we do not reach.
 
Figure~3 displays the diffusion
coefficients $D=D_{\rm eff}+D_v+D_{\rm rad}$ in the asymptotic regime
($D_{\rm rad}$ is the radiative viscosity).
In the $300$ and the $425~{\rm km \, s^{-1}}$  models,
we notice a sharp drop in $D$ around $r/R\simeq 0.95$
and $0.85$ respectively.
It corresponds to the place
where the derivative of $\O$ vanishes and thus, where no turbulence occurs.
There, the viscosity is purely radiative and
the transport of heavy elements will be strongly hindered.
One could be tempted to conclude that, for the fastest stars, there will be
two partially mixed zones, separated by a gap where very little diffusion takes
place.
Even worse, the region just outside of the convective
core is not turbulent either,
the diffusion coefficients diminishing by one or two orders of magnitude
there.
But we shall see in the next section that these conclusions are 
strongly affected
by the structural changes accompanying the evolution, in particular by the
chemical composition gradients.

Figure~3 illustrates yet another property of such rotating models, 
namely that the size of their convective core depends on rotation, with the
fastest model having the smallest core. This is in agreement with previous
studies of rotating stars (cf. e.g. Clement 1979), which predicted a lower
central temperature and a smaller convective core for a rotating star than for
a non-rotating one.

\section{Circulation and evolution}

The problem described in the previous section is rather academic.
It is merely a guide to understand some basic properties of the circulation.
However, it gives no clue as to what will happen actually to an evolving
model, and it bypasses the important role of the
$\mu$-gradients.

As is well known, the radius of early-type stars increases as they evolve on
the main-sequence. 
However, this expansion does not proceed in an homologous way, for the inner
part is shrinking instead.  
Local conservation of angular momentum would lead to
a core rotating much faster than the surface.
Since the profile of $\O$ required by
the asymptotic regime
does not change much between consecutive models, we thus 
expect a circulation to arise that transports momentum
from the center of the star to the surface.
Such a circulation will rise along the equator and descend
along the pole.
Note that in a rigidly rotating star
the circulation would be in the opposite direction (Sweet 1950).

In Fig.~4 we show the rotation profile required for this hypothetical
stationary regime at the end of the main-sequence, compared
to that on the ZAMS, for a model with an initial
surface velocity of $100~{\rm km \, s^{-1}}$.
It illustrates the need to carry angular momentum outward, since
local conservation of angular momentum would lead to a differential
rotation ratio ($\Omega _c / \Omega _s$) of order $\sim 10$.
(Note that on the TAMS, the total radius of the star has doubled
compared to the ZAMS model).

Mass loss is also an important cause of circulation. 
The constant readjustment
of the surface layers, which have to keep moving outward as
the star loses mass, involves an important circulation
in the outer part of the envelope. We thus expect outer layers
to remain well mixed.

As was mentioned before though, the main impact of evolution
is felt through the formation of a mean molecular weight gradient.
In an inhomogeneous star, the relevant quantity 
in the description of the circulation is
not $\O$ itself, but rather $\Theta - \Lambda$ (cf. Eq. 5).
As was first shown by Mestel (1953), the advection
of matter of larger mean molecular weight from the
center of the star has a choking effect on the
circulation. 
This, of course, is particularly important near the
regressing convective core. However, contrary to Mestel's
conclusions, the circulation is not completely ``killed'' by
those gradients, even though we expect it to be slowed. 
In an evolving star, the horizontal $\mu$-gradients
keep adjusting themselves to the new equilibrium
state; furthermore, due to the erosion by the horizontal turbulence, 
at least a little amount of advection is required to maintain
the horizontal inhomogeneities. 

Another critical effect of the $\mu$-gradients
is that they tend to inhibit baroclinic shear instabilities. However, as was
shown by Talon \& Zahn (1996), if the turbulence is strongly anisotropic, 
horizontal diffusion reduces the buoyancy force associated with these
gradients, just like temperature gradients are weakened
by thermal diffusion. Their prescription (\ref{TZ}) allows for vertical
mixing, provided the differential rotation is strong enough.

\subsection{Evolutionary calculations}

Complete evolutionary calculations were performed for two models,
initially rotating at $100$ and $300~{\rm km \, s^{-1}}$ (we
will later refer to these models as ``slow'' and ``fast'').
The stars are assumed to be in the asymptotic regime on the ZAMS,
thus having the rotation profiles shown in Fig.~1.
We are aware that this initial condition is arbitrary, but
the memory of the initial condition will be lost in an Eddington-Sweet
time-scale  
\beq
t_{\rm ES} = \frac{GM^2}{LR} \lp \frac{GM}{\O^2 R^3} \rp ,
\eeq
of order $10^6$ years for the considered models.
Furthermore, during this early adjustment phase, no chemical
inhomogeneities are present, and this diminishes the impact of
the initial condition on the amount of mixing.

In Fig.~5, we show the evolutionary paths followed by the two
models, plus that of two comparison models, one with a convective overshooting
of $d/H_p = 0.2$, and the other one without overshooting.
One striking feature of that diagram is that 
the mixing associated with a star initially rotating
at $100~{\rm km~s^{-1}}$ yields an evolutionary path which can
hardly be distinguished from that of a model with a small
amount of overshooting.
The initial positions of these models on the HR diagram
also illustrate the small impact of the hydrostatic corrections
on the global parameters, compared to those related to mixing.

At the end of the main-sequence, the ``slow'' model
has lost 0.09\% of its initial mass, and 1.4\% of 
its total angular momentum. The ``fast''model has lost
0.12\% and 2.3\% of its mass and momentum respectively.
Remember that we used an empirical formula for mass loss that
does not take into account the rotation rate of the star.
The difference in the mass lost only reflects different lifetimes
and effective temperatures during the evolution.

Figure~6 displays the evolution of the central mass fraction
of hydrogen, and thus also the lifetime on the main-sequence.
For the ``slow'' model, which has a velocity typical for observed
$9~M_\odot$ stars, the evolution in the HR diagram, as well
as the main-sequence lifetime, are comparable to what is obtained
with a standard model including overshooting. 

As mentioned in the Introduction, convective penetration may be
severely hindered in rotating stars, according to the recent
simulations carried out by Julien et al. (1996). Our calculations show that the
widening of the main-sequence may be due to rotational mixing, in stars
having a typical velocity ($\sim 100 ~{\rm km~s^{-1}}$).

\subsection{Role of $\mu$-gradients}

\begin{figure}[t]  
\centerline{
\psfig{figure=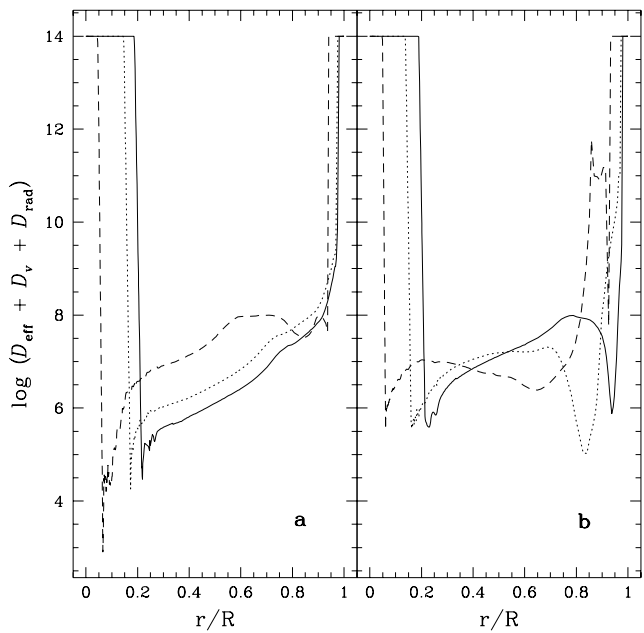,height=9.0cm}
}
\caption[]{Total diffusion coefficient and its variation
with evolution. ({\bf a}) Model with an initial surface
velocity of $100~{\rm km~s^{-1}}$. ({\bf b}) Model with an initial surface
velocity of $300~{\rm km~s^{-1}}$.
The full line refers to the youngest model, the dotted line
to the model in the middle of the main-sequence, and the dashed line to
the oldest model, at the turning point of the main-sequence.
}
\end{figure}

\begin{figure}  
\centerline{
\psfig{figure=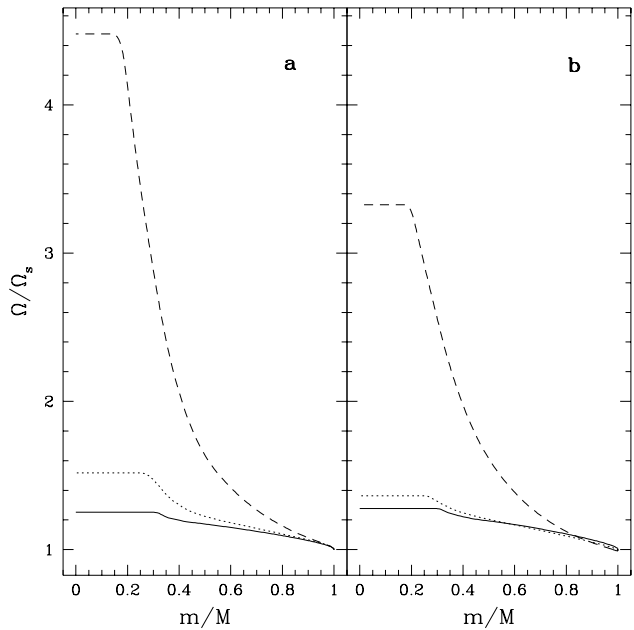,height=9.0cm}
}
\caption[]{Rotation profile and its variation with evolution.
(Line styles are defined as in Fig.~7 and refer to the same models).
}
\end{figure}

\begin{figure}  
\centerline{
\psfig{figure=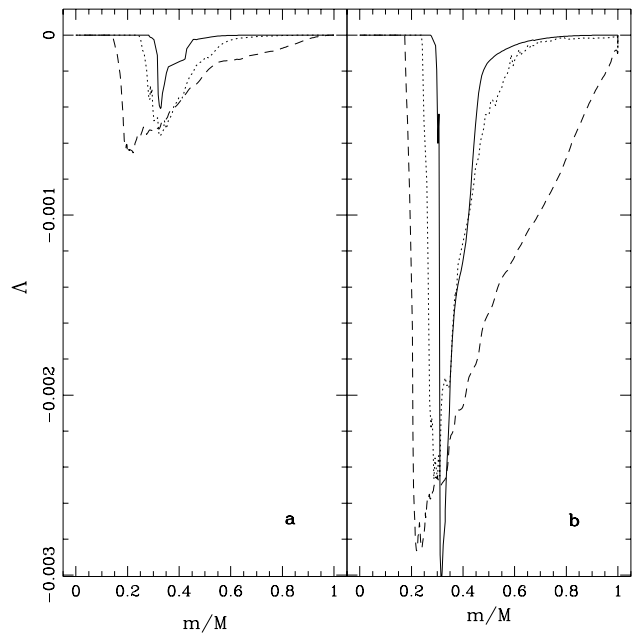,height=9.0cm}
}
\caption[]{$\Lambda = \widetilde{\mu}/\mu$ and its
variation with evolution.
(Line styles are defined as in Fig.~7 and refer to the same models).
}
\end{figure}

\begin{figure}[t]     
\centerline{
\psfig{figure=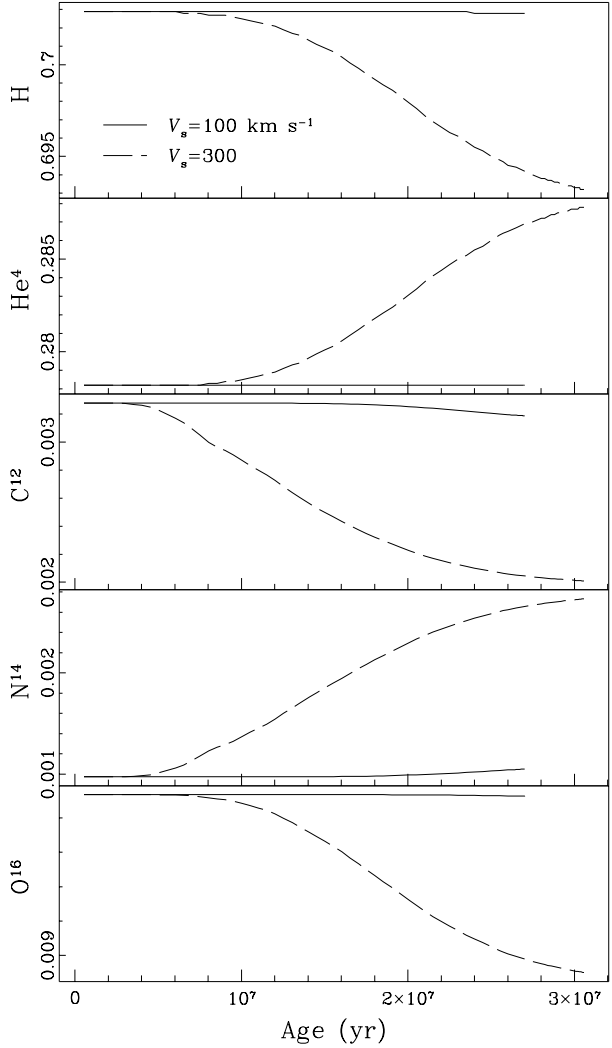,height=15.5cm}
}
\caption[]{Evolution of surface abundances.
(Line styles are defined as in Fig.~5).
}
\end{figure}

\begin{figure}[t]     
\centerline{
\psfig{figure=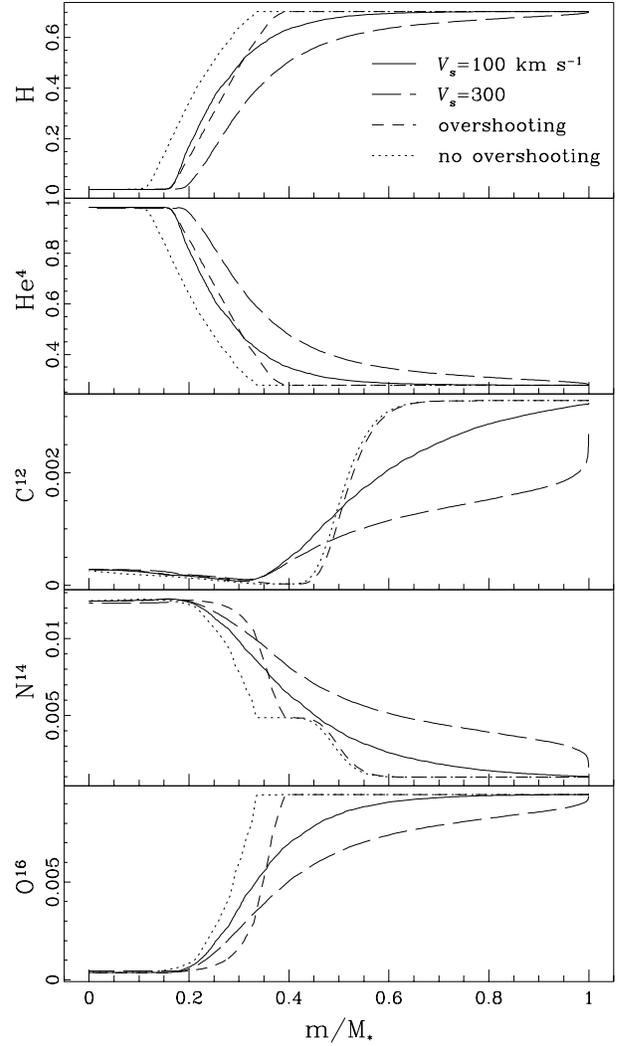,height=15.5cm}
}
\caption[]{Final abundance profiles on the TAMS of hydrogen (H), helium
(He$^4$), carbon (C$^{12}$), nitrogen (N$^{14}$) and oxygen (O$^{16}$).
(Line styles are defined as in Fig.~5).
}
\end{figure}

In the preceding section, we described the global results
obtained by solving the complete set of equations stated in \S 2, 
which include the effect of the composition gradients. 
Here we shall examine in more detail the particular role of these 
$\mu$-gradients.

In Fig.~7, we show the diffusion coefficient at three different
ages, for the two rotating models.
For the ``slow'' model, the major contribution to diffusion is 
turbulent. Close to the convective core, we note a significant
drop in the diffusion coefficient, which is due to the steep
molecular weight gradient. 
The turbulent transport is affected, but it is not
suppressed (cf. Eq.~\ref{TZ}). 

For the ``fast'' model, we note also a strong diminution of the
diffusion coefficient near the surface. It is due to the reversed
circulation cell, which grows with time. The reason for this growth
is readily found; as the star evolves off the zero-age main-sequence, its
radius grows and its breakup velocity decreases. Local conservation
of angular momentum would lead to still faster decline of
the surface velocity. However, circulation and turbulence keep
extracting angular momentum from the interior, and the actual
ratio between surface and critical velocities decreases.

The minimum of the total diffusion coefficient is thus located
where ${\rm d}\O/{\rm d}r = 0$, and it reduces there to the effective
diffusivity $D_{\rm eff}$ associated with the meridian circulation (cf.
\ref{deff}). 
That strangling will be the limiting factor in the modification of
the surface abundances of the fast models (see also Figs.~10 and 11).

Another important effect is shown on the rotation profile
within the model (cf. Fig.~8).
It is clear that those profiles are quite different from
what is required for thermal equilibrium
(see Fig.~4). The main reason for this difference is
the presence of the term in $\Lambda$ (shown in Fig.~9).
The steep profile in $\Lambda$ requires that $\Theta$ 
be also very steep there, so that the sum of the two terms
can give a profile similar to that shown in Fig.~4, which would be required
for the asymptotic regime.

There is here an interesting coupling between all the variables.
If ${\rm d}\mu/{\rm d}r$ is large, then so is $\Lambda$ (cf. Eq.~\ref{fluctc}),
and consequently, so is ${\rm d}\ln\O/{\rm d}r$. 
Such a strong differential rotation enhances
turbulent diffusion, thereby diminishing ${\rm d}\mu/{\rm d}r$.
However, as turbulent diffusion increases, so does $U_{\rm r}$, and thus
so does also $\Lambda$.

Let us stress that the behavior of $\Lambda$ depends on an adjustable parameter
$C_h$ of order unity. That parameter is also present in $D_v$, through
$D_h$ (cf. Eq.~\ref{TZ}). It is interesting to note that as $C_h$ 
increases, so does $\Lambda$, but $D_{\rm t}$ varies in the opposite
way. The net result is that the final profiles depend sensitively
on the choice of $C_h$, which has here the value 0.15.

In Fig.~10, we show the evolution of the surface
abundances of a few elements, and in Fig.~11, the final
abundance profiles.
Fast rotation is required to yield detectectable abundance
changes on the main-sequence. 
However, even for the ``slow'' model,
the post--main-sequence evolution is probably affected
by the composition changes inside the star.
Furthermore, as mass loss becomes more important with the evolution off
the main-sequence, additional changes of the surface abundances will
occur through the ``peeling'' of the outer layers.

\section{Discussion}
We completed a first set of calculations of rotational mixing
induced by meridional circulation as well as shear turbulence,
with the rotation profile evolving consistently with these
flows.

Unfortunately, very few observations permit to verify the full
validity of the results. 
Gies \& Lambert (1992) have observed galactic B stars and
measured their rotational velocity, as well as their CNO
abundances. However, they studied only slow rotators ($v~\sin i < 85\, 
{\rm km~ s}^{-1}$), which did not show strong evidence of composition changes.

One would expect binary systems to yield important
information about the precise position in the HR diagram of stars
with known masses, radii and rotation velocities. 
The best known binary in the $9~M_\odot$ mass range
is QX Car, whose masses are $9.267~\pm~0.122~M_\odot$ 
and $8.480~\pm~0.122~M_\odot$ (cf. Andersen 1991).
However, these stars are located too close to the ZAMS to give
any interesting information about rotational mixing.
Other binary systems (e.g. $\alpha$ Vir, NY Cep)
are not known with sufficient precision to be used for any
reliable comparison.

The most promising tests will be provided by the fastest rotating B-stars,
which show signs of helium enrichment. 
The difficulty will be however to sort out the effects of semi-convection
from those of rotational mixing.

Let us finally recall that a complete treatment of rotational mixing must
include the angular momentum transport by
the internal gravity waves emitted from the convective core. 
Such transport appears to be
quite important in the Sun, as pointed out recently by Kumar \& Quataert
(1996) and by Zahn, Talon \& Matias (1996). 
Since internal waves tend to restore uniform rotation, 
which would speed up the circulation, one expects  a priori more mixing to
occur. 
But the whole problem is highly non-linear, and it is difficult to
make any prediction until this wave transport has been implemented and its
impact has been evaluated.  Work is progressing in this direction.

\begin{acknowledgements}
S.T. gratefully acknowledges support from NSERC of Canada.
\end{acknowledgements}

\end{document}